\documentclass[a4paper,
              ]{jacow}
\makeatletter%
	\ifboolexpr{bool{xetex}}
	 {\renewcommand{\Gin@extensions}{.pdf,%
	                    .png,.jpg,.bmp,.pict,.tif,.psd,.mac,.sga,.tga,.gif,%
	                    .eps,.ps,%
	                    }}{}
\makeatother

\ifboolexpr{bool{xetex} or bool{luatex}} 
 {}                                      
 {\usepackage[utf8]{inputenc}}           

\usepackage[USenglish]{babel}			 

\usepackage[final]{pdfpages}
\usepackage{multirow}
\usepackage{ragged2e}
\usepackage{hyperref,graphicx,adjustbox}

\ifboolexpr{bool{jacowbiblatex}}%
 {%
  \addbibresource{jacow-test.bib}
  \addbibresource{biblatex-examples.bib}
 }{}
\begin{document}

\title{Electron Cooling Experiment for Proton Beams With Intense Space-Charge in \NoCaseChange{IOTA}}
\author{N. Banerjee\thanks{nilanjan@uchicago.edu}, M.K. Bossard, J. Brandt, Y-K. Kim, The University of Chicago, Chicago, USA \\
		B. Cathey, S. Nagaitsev\textsuperscript{1}, G. Stancari, Fermilab, Batavia, Illinois, USA \\
	\textsuperscript{1}also at The University of Chicago, Chicago, USA}
		
\maketitle
\begin{abstract}
    Electron cooling as a method of creating intense ion beams has a practical upper limit when it comes to the peak phase space density of ion beams which can be achieved in practice. We describe a new experiment to study electron cooling of 2.5 MeV protons at the intensity limit using the Integrable Optics Test Accelerator (IOTA), which is a storage ring dedicated to beam physics research at Fermilab. This system will enable the study of magnetized electron cooling of a proton beam with transverse incoherent tune shifts approaching -0.5 due to the presence of intense space-charge forces. We present an overview of the hardware design, simulations and specific experiments planned for this project.
\end{abstract}

\section{Introduction}
Electron cooling is the process of exchanging thermal energy between an ion beam and a co-propagating electron beam moving at the same average velocity. This method can be used to accumulate ions and reduce the emittance of the beam in storage rings and is especially important for future heavy-ion facilities\cite{Spiller2020,Yang2013,Kekelidze2017} and cooling in an ion collider\cite{Fedotov2020}. While its possible to generate very cold electron beams\cite{Danared1994} to cool ions, the minimum emittance of ion beams achieved through electron cooling is limited by the additional heating processes of Intra-Beam Scattering (IBS) and resonance-driven transverse heating due to space-charge tune shifts.\cite{Parkhomchuk2001} In practice, this is observed as a minimum achievable beam size in a storage ring\cite{Nagaitsev95,Steck2000} corresponding to transverse space-charge tune shifts of 0.1-0.2. We are developing an experiment to explore the interplay of electron cooling with space-charge and instabilities in a high-brightness high-intensity regime.\\

The Integrable Optics Test Accelerator (IOTA) which can store with a kinetic energy of 2.5~MeV is a suitable machine to explore electron cooling in the presence of intense space-charge forces. IOTA is a re-configurable 40~m storage ring which acts as a test facility at Fermilab dedicated to research on intense beams including the areas of Non-linear Integrable Optics (NIO), beam cooling, space-charge, instabilities and more.\cite{Antipov2016,Valishev2021} Here, we discuss the electron cooling experiment which forms a part of our electron-lens research program.\cite{Stancari2021}. We describe a setup capable of exploring the dynamics due to space-charge in the regime of large transverse tune shifts up to $\Delta \nu_{x,y} = -0.5$.\\

In the next section, we detail the design specifications of the electron cooler setup at IOTA and list some of the relevant hardware. Then we show some simulations of the expected dynamics. In the last section, we discuss our simulation results and present plans for experiments.\\

\begin{table}
    \caption{Typical Operation Parameters for Protons in IOTA}
    \label{tab:protonops}
    \centering
     \begin{adjustbox}{max width=0.95\columnwidth}
    \begin{tabular}{p{\dimexpr 0.4\linewidth-2\tabcolsep}
                    p{\dimexpr 0.25\linewidth-2\tabcolsep}
                    p{\dimexpr 0.24\linewidth-2\tabcolsep}
                    p{\dimexpr 0.11\linewidth-2\tabcolsep}}
        \hline
        \textbf{Parameter} & \multicolumn{2}{c}{\textbf{Value}} & \textbf{Unit} \\
        \hline
        Circumference ($C$) & \multicolumn{2}{c}{39.96} & m \\
        Kinetic energy ($K_b$) & \multicolumn{2}{c}{2.5} & MeV \\
        Emittances ($\epsilon_{x,y}$) & \multicolumn{2}{c}{4.3, 3.0} & $\mu$m \\
        Momentum spread ($\sigma_p/p$) & \multicolumn{2}{c}{$1.32\times10^{-3}$} & \\
        \hline
        & \textbf{Coasting} & \textbf{Bunched} &\\
        \hline
        Number of bunches & - & 4 & \\
        Bunch length ($\sigma_s$) & - & 0.79 & m \\
        Beam current ($I_b$) & 6.25 & 1.24 & mA \\
        Bunch charge ($q_b$) & 11.4 & 0.565 & nC \\
        Tune shifts ($\Delta \nu_{x,y}$) & \multicolumn{2}{c}{-0.38, -0.50} & \\
        \hline
        $\tau_{\text{IBS,x,y,z}}$ & 6.4, 4.2, 8.1 & 8.7, 6.0, 23 & s \\
        \hline
    \end{tabular}
\end{adjustbox}
\end{table}

\section{Electron Cooler Setup}
The design parameters of the electron cooler is dependent on the specific experiments to be performed and the proton beam parameters at IOTA. Table~\ref{tab:protonops} shows some baseline operation parameters along with tune shifts and emittance growth times, in both coasting beam and bunched beam configurations. At the maximum design current corresponding to a vertical space-charge tune shift of -0.5, emittance growth times due to IBS are typically less than 10~seconds. Additionally, space-charge forces also create rapid emittance growth and beam-loss in the first few hundred turns after injection. We need a strong electron cooler to mitigate these effects. At beam currents 10 times smaller then the maximum, IBS is the dominant driver of emittance growth thus limiting beam lifetime and constraining other experiments which can be performed with proton beams at IOTA. Consequently a weaker electron cooler can compensate for the heating and is valuable for all research with proton beams in IOTA.\\

We have designed two separate electron cooler configurations for IOTA: a simple cooler configuration for cooling and performing beam manipulations at relatively small beam currents where space-charge forces are weak in the proton beam and a strong cooler configuration specifically for experiments in the high-brightness regime. Table~\ref{tab:ecooler} lists the relevant parameters of the magnetized cooler configurations. The electron beam size is chosen to effectively cool at least $3\sigma$ of the transverse size of the proton beam. While the cathode temperature provides a transverse velocity distribution to the electron beam, the cyclotron motion of the electrons inside the solenoid of the cooler effectively damps their transverse temperature as seen by the ions, giving rise to stronger cooling. Since perturbations to the solenoid field can perturb the cyclotron motion of the electrons, the field flatness of the solenoid must be constrained. The maximum current of the cooler is chosen so that the velocity depression at the center of the electron beam due to space-charge forces is of the same order as the standard deviation of the velocity distribution of the ion beam. We can adjust the cooling rates of the both configurations by varying the main and cathode solenoid strengths. \\

\begin{table}
    \caption{Electron Cooler Parameters for IOTA}
    \label{tab:ecooler}
    \centering
    \begin{adjustbox}{max width=0.95\columnwidth}
    \begin{tabular}{p{\dimexpr 0.4\linewidth-2\tabcolsep}
                    p{\dimexpr 0.25\linewidth-2\tabcolsep}
                    p{\dimexpr 0.25\linewidth-2\tabcolsep}
                    p{\dimexpr 0.11\linewidth-2\tabcolsep}}
        \hline
        \textbf{Parameter} & \multicolumn{2}{c}{\textbf{Values}} & \textbf{Unit} \\
        \hline
        \multicolumn{4}{c}{\textbf{Proton parameters}} \\
        \hline
        RMS Size ($\sigma_{b,x,y}$) & \multicolumn{2}{c}{3.22, 2.71} & mm\\
        \hline
        \multicolumn{4}{c}{\textbf{Main solenoid parameters}} \\
        \hline
        Magnetic field ($B_\parallel$) & \multicolumn{2}{c}{0.1 - 0.5} & T \\
        Length ($l_\text{cooler}$) & \multicolumn{2}{c}{0.7} & m \\
        Flatness ($\langle B_\perp \rangle/B_\parallel$) & \multicolumn{2}{c}{$2\times10^{-4}$} & \\
        \hline
        \multicolumn{4}{c}{\textbf{Electron parameters}} \\
        \hline
        Kinetic energy ($K_e$) & \multicolumn{2}{c}{1.36} & keV \\
        Temporal Profile & \multicolumn{2}{c}{DC} & \\
        Transverse Profile & \multicolumn{2}{c}{Flat} & \\
        Source temp. ($T_\text{cath}$) & \multicolumn{2}{c}{1400} & K \\
        \hline
        Current ($I_e$) & 1 & 40 & mA \\
        Radius ($a$) & 20 & 12 & mm \\
        $\tau_\text{cool,x,y,s}$ & 37, 33, 18 & 3.4, 3.2, 1.7 & s\\
        \hline
    \end{tabular}
\end{adjustbox}
\end{table}

\begin{figure}
    \centering
    \includegraphics[width=.95\columnwidth]{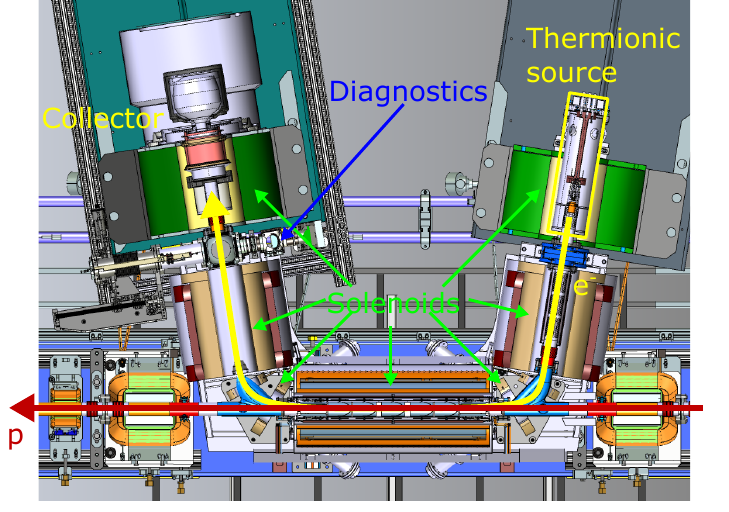}
    \caption{Electron lens setup for IOTA, which will also act as the electron cooler.}
    \label{fig:ecooler}
\end{figure}

Figure~\ref{fig:ecooler} depicts a model of the electron cooler setup showing the thermionic source, collector and all the solenoids which keep the electron beam magnetized throughout the transport line. While the strengths of the source and the main solenoid can be adjusted to provide a dynamic range of a factor of 10 in terms of electron current density, the baseline design assumes that both the source and the main solenoid are set to 0.1~T. Diagnostics for the electron beam include a Faraday cup to measure current and a scintillating screen to image the transverse profile at the entrance to the collector. Proton beam diagnostics include beam position monitors and a DC current transformer. Additionally, radiative recombination of the protons and the electrons inside the cooler allow for a non-destructive but slow (compared to cooling time) diagnostic of the equilibrium transverse profile of the protons by using a micro-channel plate detector and associated imaging system\cite{Tranquille2018}. The measurement of proton beam lifetime and equilibrium transverse profiles enable the realization of a variety of electron cooling experiments at IOTA.\\

\begin{figure}
    \centering
    \includegraphics[width=\columnwidth]{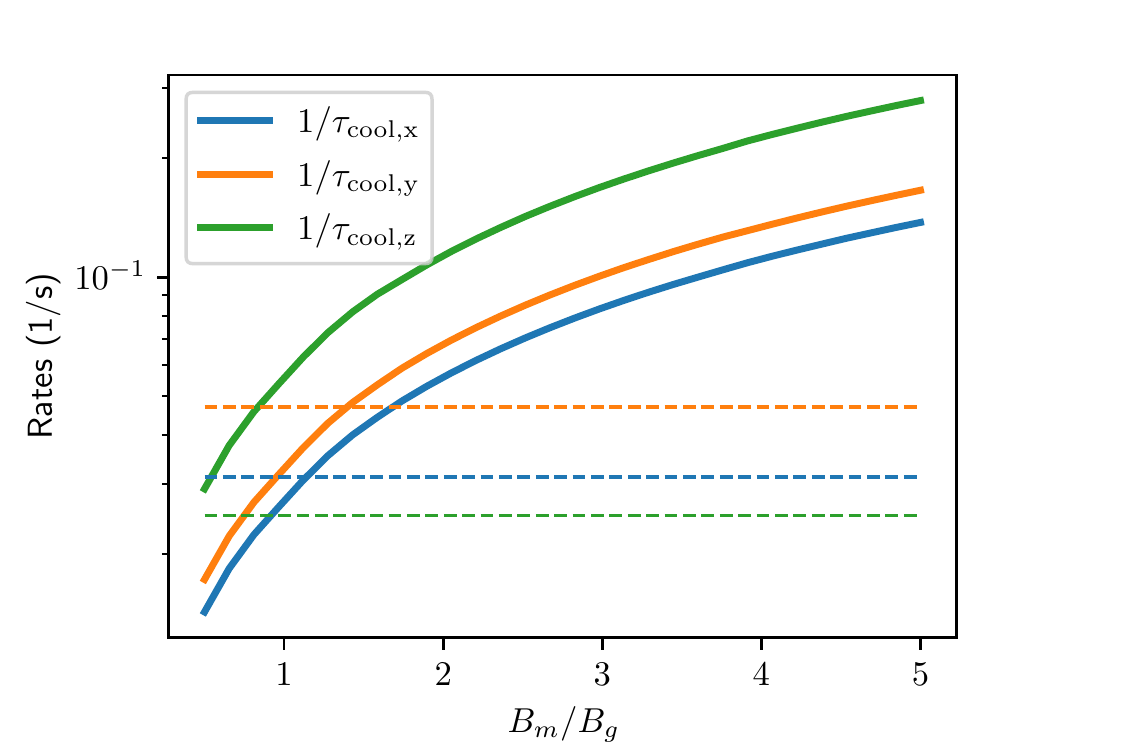}
    \caption{Cooling rates for the simple cooler configuration as a function of the ratio of the main and cathode solenoid fields. Dashed lines represent emittance growth rates due to IBS for a 1.25~mA coasting proton beam.}
    \label{fig:simplecooler}
\end{figure}

\begin{figure*}
    \centering
    \includegraphics[width=0.95\linewidth]{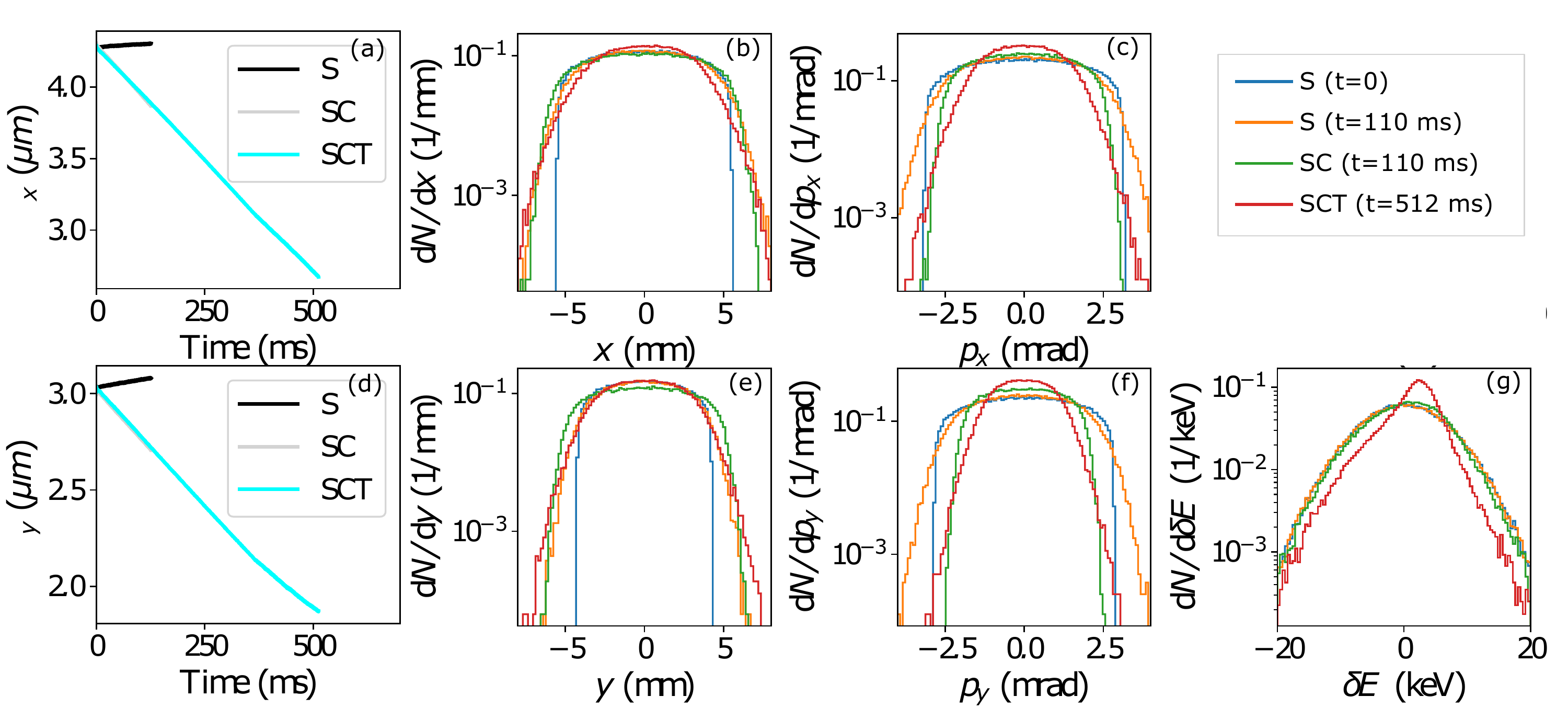}
    \caption{Electron cooling simulation results for IOTA with transverse space charge using the \texttt{eccoler} extension in PyORBIT. Panels (a) and (d) show the evolution of transverse emittance as a function of time for various runs listed in the text. Other panels show snapshots of particle distribution on various axes of phase space.}
    \label{fig:results}
\end{figure*}

\section{Simulations}
The simple cooler configuration will be used for the low current regime, where space-charge tune shifts of the ions are less than 0.1 and IBS is the major driver of emittance growth. We used the electron cooling simulation code JSPEC\cite{Zhang2021} to estimate the cooling times as a function of beam parameters. Figure~\ref{fig:simplecooler} shows the dynamic range of cooling times achievable by adjusting the solenoid fields at the cooler ($B_m$) and at the cathode ($B_g$). The dashed lines represent the emittance growth rates due to IBS for a proton beam current of 1.25~mA. Depending on the ion current, we will adjust the cooling rates during operations to compensate for IBS and increase the beam lifetime.\\

The strong cooler configuration is designed to cool proton beams where the transverse space-charge tune shift is more than 0.1. We implemented an \texttt{ecooler} extension\cite{Banerjee2021} using the Parkhomchuk model\cite{Parkhomchuk2000} to the particle-in-cell space-charge code PyORBIT\cite{Shishlo2015}. For a coasting beam at the design current of 6.25~mA, the space-charge forces dominate the dynamics over electron cooling with the strong cooler parameters listed in Table~\ref{tab:ecooler} and the simulations predict long term emittance growth. Simulations starting with a tune shift of $\Delta \nu_y = -0.1$ corresponding to a coasting beam current of 1.25~mA show interesting dynamics and some results are depicted in Figure~\ref{fig:results}. The transverse emittance shown in panels (a) and (d) grow for the simulation which only includes space-charge labelled S, while the emittance decays for run SC with both space-charge forces and cooling. To speed up the cooling simulation, we introduce an additional run SCT where cooling force is scaled by a factor of 10 for the first 20000 turns ($\approx 36$~ms) and then switched back to design values for the rest of the simulation. While plotting, the time axis is suitably scaled to account for the stronger cooling. The 1D histograms of particle distribution in $x$, $p_x$, $y$ and $p_y$ at $t=110$~ms in panels (b), (c), (e) and (f) respectively indicate that while space-charge forces lead to the diffusion of particles from the core of the beam into higher amplitudes, electron cooling concentrates more particles into the core of the distributions. Further, panel (g) shows clear evidence of cooling in the longitudinal space.\\

\section{Discussion and Outlook}
We discuss the design of a new electron cooler experiment using 2.5~MeV protons at IOTA. A simple cooler configuration operating at low currents will be used to compensate for emittance growth in beams with weak space-charge and perform beam manipulations. Additionally, a strong cooler configuration will be used to analyze the interplay of electron cooling and space-charge forces. We present the design parameters of both configurations and simulations of their performance. The strong cooler configuration along with suitable beam diagnostics will allow us to probe both incoherent dynamics due to space-charge and coherent dynamics such as instabilities.\cite{Burov2018} Ongoing work includes analyzing the influence of radial beams profiles for the electron beam and design of the thermionic electron sources to produce them, incorporating IBS in the PyORBIT simulations and analyzing the heating mechanisms which appears in electron coolers at high electron currents.\\

\section*{Acknowledgement}
We would like to thank A. Valishev, A. Romanov and V. Lebedev for discussions on proton operations in IOTA. This manuscript has been authored by Fermi Research Alliance, LLC under Contract No.~DE-AC02-07CH11359 with the U.S.\ Department of Energy, Office of Science, Office of High Energy Physics. This research is also supported by the University of Chicago.

\end{document}